\documentclass[prl,showpacs,reprint,twocolumn]{revtex4}
\usepackage{amsmath}
\usepackage{braket}
\newcommand{\tr}[1]{\mathrm{Tr}[#1]}
\usepackage{bm}
\usepackage[dvipdfm]{graphicx}

\begin{document}
\title{Eigenstate Randomization Hypothesis:\\
Why Does the Long-Time Average Equal the Microcanonical Average?}
\date{\today}
\author{Tatsuhiko N. Ikeda$^1$, Yu Watanabe$^1$, and Masahito Ueda$^{1,2}$}
\affiliation{$^1$Department of Physics, University of Tokyo, 7-3-1 Hongo, Bunkyo-ku, Tokyo 113-0033, Japan\\
$^2$ERATO Macroscopic Quantum Control Project, JST, 2-11-16 Hongo, Bunkyo-ku, Tokyo 113-0033, Japan}

\begin{abstract}
We derive an upper bound on the difference
between the long-time average and the microcanonical ensemble average of observables
in isolated quantum systems.
We propose, numerically verify, and analytically support a new hypothesis, eigenstate randomization hypothesis (ERH),
which implies that in the energy eigenbasis the diagonal elements of observables
fluctuate randomly.
We show that ERH includes eigenstate thermalization hypothesis (ETH)
and makes the aforementioned bound vanishingly small.
Moreover, ERH is applicable to integrable systems for which ETH breaks down.
We argue that the range of the validity of ERH determines that of the microcanonical description. 
\end{abstract}
\pacs{05.30.-d, 03.65.-w}
\maketitle

Statistical mechanics is among the most successful fields of physics and has an extremely diverse range of applications.
However, why statistical mechanics works so well has only recently begun to be understood.
To answer this question,
we must address the problems of equilibration and thermalization.
In other words, does a given time-dependent physical quantity relax to a certain constant value?
If so, can the equilibrated value be calculated by invoking statistical ensembles?
If a system interacts with energy-exchanging reservoirs,
the canonical ensemble is known to be useful for describing the system.
Several works~\cite{tasaki,ct,popescu} have exemplified
that entanglement between the system and reservoirs plays a critical role
in justifying the canonical ensemble.

When the system is isolated, or if there is no reservoir,
the microcanonical ensemble is useful for describing the system.
Reimann~\cite{reimann} addressed the issue of equilibration and
showed that time-dependent expectation values of observables
take on their long-time averages (LTA) most of the time.
However, to justify the method of the microcanonical ensemble,
we have yet to address the problem of thermalization,
and understand why LTA is equal to the microcanonical ensemble average (MEA).

The problem of thermalization has been addressed mainly by two approaches.
One is based on eigenstate thermalization hypothesis (ETH)~\cite{deutsch,srednicki,rigol},
which states that diagonal elements of observables in the eigenenergy basis take on a constant value in the thermodynamic limit.
If ETH holds, then LTA=MEA is satisfied.
However, no proof of ETH has been given and
numerical tests are hampered due to computational complexity.
Moreover, ETH is known to break down in integrable systems~\cite{rigol}.
The other approach is to assume a virtual probability distribution~\cite{fremoro}.
This approach assumes that
the initial state of the system is generated according to the virtual probability distribution.
In this case, LTA=MEA holds with a high probability.
However, in reality, the initial state is not generated in such a manner,
and therefore the physical ground of this approach is unclear.

In this Letter, we show LTA=MEA by resolving the problem of thermalization.
Firstly, we show an inequality,
which gives an upper bound on the difference between LTA and MEA.
Secondly, as a sufficient condition for LTA=MEA, we derive an eigenstate randomization hypothesis (ERH),
which implies that the diagonal elements of observables
behave randomly in the eigenenergy basis.
We show numerically that ERH holds for few-body observables whether or not ETH holds.
Moreover, we show that 
the microcanonical description is justified not just for expectation values
but also for higher moments
and the extent to which ERH holds delimits the range of validity beyond which the microcanonical description breaks down.
Lastly, we show analytically that ERH holds for almost all observables.

We consider an isolated many-body quantum system, which is described by Hamiltonian $H$,
and a few-body observable $A$.
Here by few, we mean a number much less than the number of particles in the system.
Let the eigenvalues and eigenvectors of $H$ be $E_1 \leq E_2 \leq \dots \leq E_{N_0} $,
and $\ket{E_1},\ket{E_2},\dots , \ket{E_{N_0}}$, respectively.
We assume that
for $E_{i_1}\neq E_{i_2}$ and $E_{j_1}\neq E_{j_2}$,
$E_{i_1}-E_{i_2}=E_{j_1}-E_{j_2}$ if and only if $E_{i_1}=E_{j_1}$ and $E_{i_2}= E_{j_2}$.
This condition is necessary for the many-body quantum system to equilibrate
according to the theorem found by Reimann and its generalization~\cite{reimann,comment}.
If $H$ has some degeneracies,
we choose $\{ \ket{E_i}\}_{i=1}^{N_0}$ to diagonalize $A$ in each degenerate subspace.
The initial state $\ket{\psi}$ is expanded in terms of energy eigenstates as $\ket{\psi} = \sum_i C_i\ket{E_i}$.
We assume that the coefficients $\{C_i\}_{i=1}^{N_0}$ satisfy the following condition.
There exists a macroscopically small energy width $\varDelta E$
such that $C_i \neq 0$ only if  $E_i \in [E-\varDelta E,E+\varDelta E]$,
where $E\equiv \braket{\psi |H| \psi}$.
We assume the number of eigenenergies in the interval $[E-\varDelta E,E+\varDelta E]$
is much greater than one,
yet $\Delta E$ is much smaller than the resolution limit of an experimental apparatus.

Now, LTA of the observable $A$ is defined as
$\text{LTA}_{\psi}(A) \equiv \lim_{T\rightarrow \infty}T^{-1}\int_0^T \braket{\psi(t)|A|\psi(t)}\text{d}t$.
Based on the assumptions described above, we can show that
\begin{equation}
\label{lta}
\text{LTA}_{\psi}(A) = \sum_i |C_i|^2 A_i,
\end{equation}
where $A_i \equiv \braket{E_i|A|E_i}$.
Given an energy window $\delta$ $(\gtrsim \varDelta E)$~\cite{window},
MEA of the observable $A$ is defined as
\begin{equation}
\text{MEA}_{\delta}(A) \equiv N(E,\delta)^{-1}\sum_{ |E_i -E|\leq \delta }A_i,
\end{equation}
where $N(E,\delta)$ is the number of energy levels that satisfy ${ |E_i -E|\leq \delta }$
and will be denoted simply as $N$.
Note that $N$ can be as large as $N\sim 10^{10^{23}}$ for macroscopic systems.
For the sake of convenience, we relabel $i$ so that
$\{ E_i \}_{i=0}^{N-1}$ appear in $[E-\delta,E+\delta]$ 
in ascending order.
As shown later, the inequality
\begin{align}
\label{ineq}
\left| \text{LTA}_{\psi}(A) - \text{MEA}_{\delta}(A) \right|  \leq \sigma_A D \left( \frac{\sigma(m)}{\sigma_A}+\frac{m}{N} \right),
\end{align}
which gives an upper bound on the difference between LTA and MEA,
holds for any positive integer $m$ $(\le N)$.
Here, 
$\sigma_A^2$ is the variance of the sequence
$\{ A_i \}_{i=0}^{N-1}$,
$\sigma_A^2\equiv N^{-1}\sum_{i=0}^{N-1}(A_i-\overline{A})^2$,
where $\overline{A}\equiv N^{-1}\sum_{i=0}^{N-1}A_i$,
$\sigma(m)^2$ is the variance of coarse-grained $A$,
which is defined below Eqs. \eqref{trans1} and \eqref{trans2},
and $D$ is a measure of the smoothness of the wave function such that $| \, |C_{i+1}|^2-|C_i|^2 \, | \leq D/N^2$.

Now, we define our coarse-graining procedure and $\sigma(m)$.
Given a positive integer $m$, we divide $N$ by $m$
and define the quotient $M$ and the remainder $n$ as follows:
$M\equiv \lceil N/m \rceil$, or $M$ is the smallest integer not less than $N/m$, and $n \equiv N-(M-1)m$.
We group the elements of the sequence $\{ A_i\}_{i=0}^{N-1}$ $m$ by $m$
and define another sequence $\{ B_k\}_{k=0}^{M-1}$:
\begin{align}
\label{trans1}
&B_k \equiv \frac{1}{m}\sum_{\alpha=0}^{m-1}  A_{km+\alpha} \quad (k=0,1,\dots,M-2),\\
\label{trans2}
&B_{M-1} \equiv \frac{1}{n}\sum_{\alpha=0}^{n-1} A_{(M-1)m+\alpha}.
\end{align}
We call $\{B_k\}_{k=0}^{M-1}$ the $m$-th coarse-grained sequence of $\{ A_i\}_{i=0}^{N-1}$.
The variance of the new sequence is
$\sigma(m)^2\equiv M^{-1}\sum_{k=0}^{M-1}(B_k-\overline{B})^2$,
where $\overline{B}\equiv M^{-1}\sum_{k=0}^{M-1}B_k$.
Then,
the first coarse-grained sequence is $\{ A_i\}_{i=0}^{N-1}$ itself,
so $\sigma(1)=\sigma_A$.
Note that if the sequence $\{ A_i\}_{i=0}^{N-1}$ behaves randomly,
fluctuations of its $m$-th coarse-grained sequence are greatly suppressed
and its variance $\sigma(m)^2$ is expected to decrease as $m$ increases.

Let us consider the meaning of the inequality~\eqref{ineq},
which consists of the summation of two terms:
$\sigma(m)/\sigma_A$ and $m/N$.
As we make further coarse graining and increase $m$,
the first term $\sigma(m)/\sigma_A$ decreases
depending on the randomness of the matrix elements of the observable $A$.
On the other hand, the second term increases and
its magnitude is determined by the smoothness of the weights of the wave function.
So, in making the upper bound of \eqref{ineq} minimal,
we have a trade-off between
the randomness of the observable and the smoothness of the wave function.


Now, let us consider sufficient conditions for LTA=MEA.
If ETH, which asserts that the first factor $\sigma_A$ vanishes in the thermodynamic limit,
holds, then LTA=MEA holds.
However, there exists another sufficient condition that
$\sigma(m)$ decreases rapidly as $m$ increases.
So, we come to ERH,
which asserts $\sigma(m) = \sigma_A m^{-\gamma}$,
where $\gamma$ is a positive parameter.
As explained before, this implies that the sequence $\{ A_i\}_{i=0}^{N-1}$ behaves randomly.
Note that, if ETH holds, ERH automatically holds,
since $\sigma (m)=0$ for every $m$.
In this sense, ERH includes ETH as a special case.
If ERH holds,
by minimizing the upper bound of \eqref{ineq} with respect to $m$,
we get
\begin{equation}
\left| \text{LTA}_{\psi}(A) - \text{MEA}_{\delta}(A) \right| \leq \sigma_A D' N^{-\frac{\gamma}{\gamma+1}},
\end{equation}
where $D' = \gamma^{1/(\gamma+1)}(1+\gamma^{-1})D$.
If the weights $\{|C_i|^2\}_i$ are smooth and $D=O(N^0)$,
the right-hand side is vanishingly small.
Note that the requirement on the smoothness of weights
is much weaker and actually $D=o(N^{-\gamma/(\gamma+1)})$.

To show numerical evidence for the behavior
$\sigma(m)= \sigma_A m^{-\gamma}$ $(\gamma >0)$,
we adopt a hard-core Bose-Hubbard model:
\begin{equation}
H = -J\sum_{(i,j)}(b_i^\dagger b_j+\text{H.c.})+U\sum_{(i,j)}n_in_j,
\end{equation}
where $(i,j)$ stands for the nearest neighbors.
The operators $b_i^\dagger$ and $b_i$ are the creation and annihilation operators of site $i$, respectively,
and satisfy 
$[b_i,b_j]=[b_i^\dagger,b_j^\dagger]=[b_i,b_j^\dagger]=0$ for $i\neq j$,
$\{b_i,b_i\}=\{b_i^\dagger,b_i^\dagger \}=0$, and $\{b_i,b_i^\dagger\}=1$.
The operator $n_i \equiv b_i^\dagger b_i$ is the number operator of site $i$.
In the following we measure energy in units of $J$ and take $U=0.1$.
We take 21 sites and arrange them on a 3 by 7 rectanglar lattice.
We denote the lattice constant as $d$,
and each lattice point as $\bm{x}_i = (l_xd,l_yd)$ $(l_x=0,1,\dots,6$ and $l_y=0,1,2)$.
Since the Hamiltonian conserves the total number of particles $N_{\text{p}}$,
the dynamics proceeds in the subspace with a definite value of $N_{\text{p}}$.
Here, we choose $N_{\text{p}}=5$
due to the limitation of computational complexity.
The dimension of the subspace is $\binom{21}{5}\sim2\times 10^4$.
These settings are to be compared with those by Rigol \textit{et al.}~\cite{rigol},
who studied ETH.

The observable $A$ we choose here is the quasi momentum distribution
\begin{equation}
n(k_x,k_y) \equiv \frac{1}{L_xL_y}\sum_{i,j}\text{e}^{\text{i}(\bm{x}_i-\bm{x}_j)\cdot \bm{p}}b_i^\dagger b_j,
\end{equation}
where $L_x$ and $L_y$ are the sides of the rectangle ($L_x=7d$ and $L_y=3d$)
and $\bm{p}=2\pi (k_x/L_x,k_y/L_y)$ $(k_x=0,1,\dots,6$ and $k_y=0,1,2)$
is the wave vector.
Here, we have set $\hbar = 1$.
The average energy and energy width of the initial state are chosen as $E=0$ and $\varDelta E=0.1$.
The energy width is chosen so that the central value of the sequence $\{ A_i \}_{i=0}^{N-1}$ does not vary
in the region $[E-\varDelta E,E+\varDelta E]$.

\begin{figure}
	 \includegraphics[width=8cm,clip]{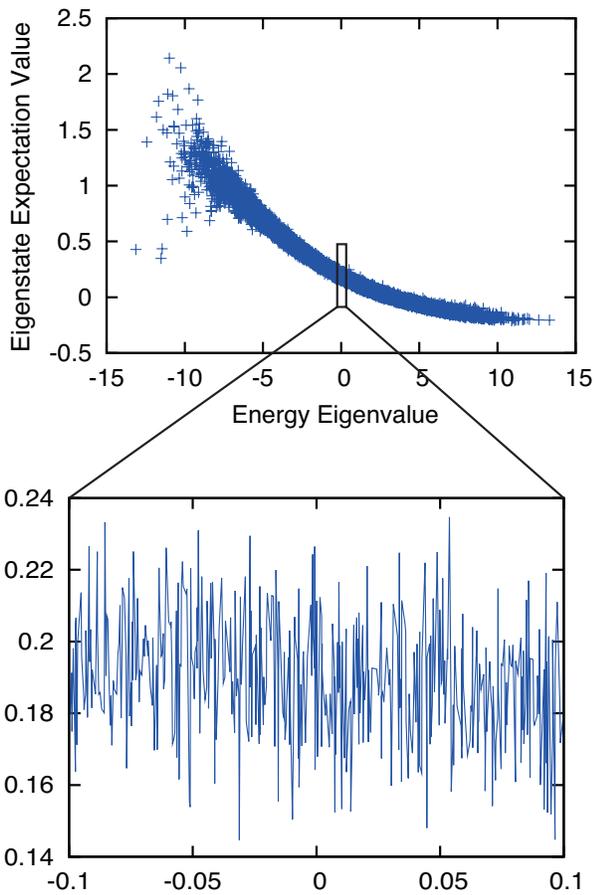}
	 \caption{
	 (Color online)
	 Energy eigenvalue $E_i$ (in the units of hopping parameter $J$)
	 versus the corresponding eigenstate expectation value of the observable $n(0,0)$
	 ($\braket{E_i|n(0,0)|E_i}$).
	 Globally $\braket{E_i|n(0,0)|E_i}$ varies smoothly (upper figure),
	 whereas locally it oscillates randomly around its local median (lower figure).
	 This local behavior reflects the fact that
	 even if two adjacent energy eigenvalues are very close,
	 the corresponding many-body wave functions are quite different.}
	 \label{diag}
	 \end{figure}

\begin{figure}
	 \includegraphics[width=8cm,clip]{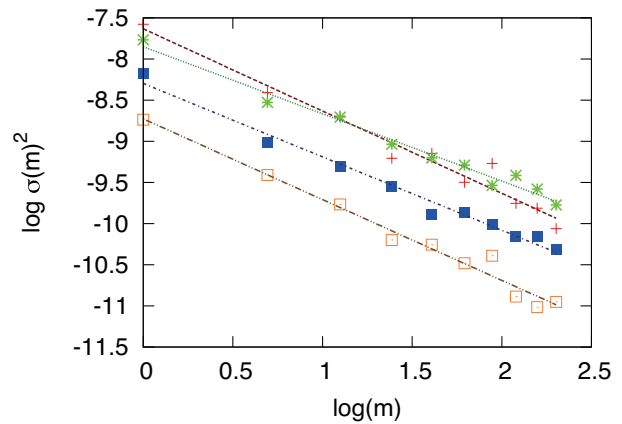}
	 \caption{
	 (Color online)
	 The variance of the $m$-th coarse-grained sequence $\sigma (m)^2$ plotted
	 for four momentum distributions $n(k_x,k_y)$ for a non-integrable system:
	 $(k_x,k_y)=(0,0)$ (cross), $(0,1)$ (asterisk),
	 $(1,0)$ (filled square), $(1,1)$ (open square).
	 The moment $\sigma(m)$ decays as $\sigma(m)=\sigma(1)m^{-\gamma}$.
	 The exponent $\gamma$ of the four observables are
	 $\gamma = 0.50, 0.41, 0.45$ and $0.49$, respectively.
	 The decay in the moment occurs because
	 the original sequence (lower figure of Fig.~\ref{diag}) behaves very randomly.	
	 }
	 \label{smdrect}
	 \end{figure}

The relation between energy eigenvalue $E_i$ and 
the corresponding eigenstate expectation value $A_i$
is illustrated in the lower figure of Fig.~\ref{diag}.
If we focus on a small energy width,
$A_i$ behaves randomly around a certain value as expected.
The decay of the variance $\sigma(m) = \sigma_A m^{-\gamma}$ caused by this random behavior
is illustrated for four choices of $n(k_x,k_y)$ in Fig.~\ref{smdrect}, where $\gamma$ is found to be between 0.41 and 0.50.

\begin{figure}
	 \includegraphics[width=8cm,clip]{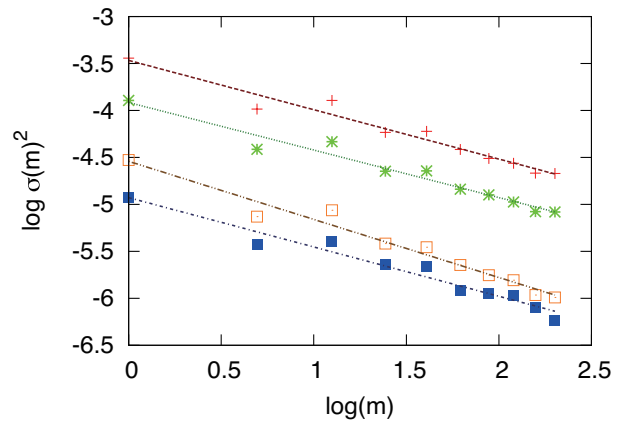}
	 \caption{
	 (Color online)
	 The variance of the $m$-th coarse-grained sequence $\sigma (m)^2$ plotted
	 for four momentum distributions $n(m)$ in the integrable system:
	 $k=0$ (cross), $1$ (asterisk),
	 $2$ (filled square), $3$ (open square).
	 The exponent $\gamma$ is found to be
	 $\gamma = 0.26$ $(k=0)$, 0.25 $(k=1)$, 0.26 $(k=2)$ and $0.31$ $(k=3)$.
	 }
	 \label{smdring}
	 \end{figure}

Changing the configuration of the system from the rectangle to the one-dimensional loop,
we have performed the same analysis.
By this change, the system becomes integrable~\cite{rigol2}.
In this case lattice points are denoted as $x_i = l_id$ $(l_i=0,1,\dots,20)$
and the quasi momentum distribution is
\begin{equation}
n(k) \equiv \frac{1}{L}\sum_{i}\text{e}^{\text{i}(x_i-x_j)p}b_i^\dagger b_j,
\end{equation}
where $p=2\pi k/L$ and $L=20d$.
The result is also illustrated for four choices of $n(k)$ in Fig.~\ref{smdring}, where
we have chosen $E=0$ and $\varDelta E=0.2$.
Although ETH is known to break down in this system~\cite{rigol},
$\sigma(m)$ decays as in the nonintegrable case.
Therefore, ERH holds and LTA=MEA is satisfied.
In this case we find $\gamma$ to be between 0.25 and 0.31.
Since $\{ A_i\}_{i=0}^{N-1}$ oscillates very drastically,
ETH breaks down,
but ERH holds because of the rapid oscillations.
Thus, even if ETH breaks down, LTA=MEA still holds.
By these analyses,
we find numerically that the moment decays as $\sigma(m)=\sigma_Am^{-\gamma} (\gamma >0)$
whether the system is integrable or not.

We have numerically showed that ERH holds for few-body observables.
However, ERH does not hold for all observables.
The projection operator onto the $i$-th energy eigenvector $P_i \equiv \ket{E_i}\bra{E_i}$ violates ERH
but this is a full $N_{\text{p}}$-body observable.
Let us assume ERH holds for less than $K_\text{max}$-body observables.
Let $A$ be a $K$-body observable,
then $A^n$ $(n=1,2,\dots)$ is an $nK$-body observable and
$\text{LTA}_{\psi}(A^n) \sim \text{MEA}_{\varDelta E}(A^n)$ $(n=1,2,\dots,\lfloor K_\text{max}/K\rfloor)$.
Our results demonstrate that the microcanonical description is valid not only for
expectation values but also for such higher moments.

In addition to the numerical evidences,
we can show ERH holds for almost all observables.
Precisely speaking,
if we consider the uniform measure on the set of $N\times N$ Hermitian matrices
whose trace norm is $gN^2$, $\mathcal{A}\equiv \{ A\,\, |\,\, \tr{A^\dagger A} = gN^2, A^\dagger = A\}$,
ERH holds for almost all matrices,
where $g$ is a constant of the order of unity and
the factor $N^2$ implies that every element of $A$ does not vanish or diverge as $N$ increases.
The set of matrices $\mathcal{A}$ can be considered as
an $(N^2-1)$-dimensional sphere whose radius is $\sqrt{gN^2}$.
Note that \eqref{trans1} and \eqref{trans2} are linear transformations
such that they are written in vector representation as $\bm{B}=C(m)\bm{A}$,
where $C(m)$ is the $M \times N$ matrix.
The variance $\sigma(m)^2$ is calculated to be $\sigma(m)^2=M^{-1}\bm{A}^\mathrm{T} C(m)^\mathrm{T}
(I-M^{-1}\bm{1}_M)C(m)\bm{A}\equiv
\bm{A}^\mathrm{T} W(m)\bm{A}$,
where $\bm{1}_M$ is the $M\times M$ matrix whose elements are all equal to unity.
Then,   
$\langle \sigma(m)^2\rangle
=g\tr{W(m)}$,
and
$\langle \sigma(m)^4\rangle
= g^2(1+2/N^2)^{-1}\{ \tr{W(m)}^2 +2\tr{W(m)^2}\}$,
where $\langle \dots \rangle$ means that the average is to be taken over the $(N^2-1)$-dimensional sphere.
For the sake of simplicity, here we consider the case in which
$N$ is divisible by $m$, or $n=m$ and $N=Mm$.
The eigenvalues of $W(m)$ are
$0$ and $1/N$, and 
their degrees of degeneracy are $N-N/m+1$ and $N/m-1$, respectively.
When $m\ll N$, $\langle \sigma(m)^2\rangle \sim gm^{-1}$
and $\sqrt{\langle  \sigma(m)^4 \rangle - \langle  \sigma(m)^2 \rangle^2}/\langle  \sigma(m)^2  \rangle
\sim \sqrt{2m/N} \ll 1$.
These show that ERH holds for almost all observable $A$
with $\gamma=1/2$.

Finally, we prove \eqref{ineq}.
We denote $q_i\equiv|C_i|^2$,
Then, the $m$-th coarse-grained sequences of $\{ q_i \}_{i=0}^{N-1}$,
$\{ r_k \}_{k=0}^{M-1}$, is defined as
$\bm{r}\equiv C(m)\bm{q}$.
First, we use the triangular inequality and Schwarz's inequality to show
$| \text{LTA}_{\psi}(A) - \text{MEA}_{\varDelta E}(A) |
\leq | \sum_{k,\alpha} \varDelta A_{km+\alpha}(r_k - 1/N)|
+| \sum_{k,\alpha} \varDelta A_{km+\alpha}(q_{km+\alpha} - r_k)|
\leq \sigma(m) [ (N+m ) (Q(m)-1/N)]^{1/2}
+ \sigma_A [N(Q-Q(m))]^{1/2}$,
where $Q(m)\equiv \sum_{k,\alpha} r_k^2$.
According to Cauchy-Schwarz's inequality,
$1/N\leq Q(m)\leq Q(1)$.
Moreover, by summing both sides of $|q_i-q_j|\leq |i-j|D/N^2$
over $i$ and $j$,
we have $2(NQ(1)-1)\leq D^2$.
From $| q_{km+\alpha}-r_k |\leq Dm/2N^2$,
$Q-Q(m)\leq D^2m^2/N^3$.
This proves \eqref{ineq}.

In conclusion, we have proposed a new scenario (ERH with \eqref{ineq}) under which LTA=MEA holds
in isolated many-body quantum systems.
Together with the work by Reimann~\cite{reimann}
and its generalization~\cite{comment},
this scenario explains why the method of the microcanonical ensemble
gives the correct description of quantum many-body systems.
Our scenario is valid whether or not ETH 
holds.
We verify ERH numerically in two typical cases of the hard-core Bose-Hubbard model.
It is expected that ERH holds for other configurations and observables of this model.
In fact, it is suggested that ERH holds in chaotic systems~\cite{peres1,peres2}.
Moreover, we have analytically shown that ERH holds for almost all observables.
Once we find that ERH holds for up to $K_{\rm max}$-body observables,
we also know that it sets the range of validity of the microcanonical description
and this knowledge can be used as an operational criterion of macroscopic observables.

Fruitful discussions with H. Tasaki are gratefully acknowledged.
This work was supported by KAKENHI 22340114, 
a Grant-in-Aid for Scientific Research on Innovation Areas
"Topological Quantum Phenomena"(KAKENHI 22103005),
the Global COE Program "the Physical Sciences Frontier,"
and the Photon Frontier Network Program,
from MEXT of Japan.
Y.W. acknowledges support from JSPS (Grant No. 216681).


\begin{thebibliography}{99}
\bibitem{comment}
	Reimann gave an upper bound on the long-time variance,
	$\lim_{T\rightarrow \infty}T^{-1}\int_0^T [\braket{\psi(t)| A | \psi(t)}-\text{LTA}_{\psi}(A)]^2\text{d}t$.
	While he assumed that $H$ has no degeneracy,
	we can generalize his result to the case in which $H$ has some degeneracies.
	In this case, the generalized upper bound includes an extra factor $2K-1$,
	where $K$ is the maximum number of degeneracies.
	The detailed discussion of this finding will be reported elsewhere.
\bibitem{window}
	When $\delta \lesssim \varDelta E$,
	an $O(N^{-1})$ correction is added to the right-hand side of \eqref{ineq}.
	However, this term does not qualitatively change our result.
\end{thebibliography}
\end{document}